\documentclass[11pt, ]{article}
\usepackage{cite}

\usepackage{graphicx}
\usepackage{array}

\title{50 YEARS OF NEUTRINO PHYSICS%
\thanks{Lecture presented at the L Cracow School of Theoretical Physics
``Particle Physics at the Dawn of the LHC'', Zakopane, Poland, June
9--19, 2010.}}
\author{MAREK ZRALEK\vspace{2mm}\\
Department of Field Theory and Particle Physics\\ 
Institute of Physics, University of Silesia\\ Uniwersytecka 4, 40-007
Katowice, Poland}
\date{December 10, 2010}
\begin{document}
\maketitle
\vspace{-10mm}
\begin{abstract}
Some important topics from history of neutrino physics  over the last fifty years  are
discussed. History of neutrinos is older, at 4th December 2010 it will be  eightieth
anniversary of the {\it neutrino birth}. In that day W.~Pauli wrote the famous letter to
participants of the physics conference at Tubingen with the suggestion that ``there
could exist in the nuclei electrically neutral particle''. We will
concentrate mostly on the
50 years of neutrino history just to show the long tradition of the Zakopane
Theoretical School.
\end{abstract}


\section{Introduction --- neutrinos before 1960}
\vspace{2mm}
The origins of neutrinos are related to the discovery of $\beta$ decay of nuclei at the late
of 19th century. Observation of the particles after decay pointed to a lack of 
conservation of energy and momentum in the observed process. Missing energy and momentum
could be explained by the existence of some new particle or, as Niels Bohr suggested, having
the Quantum Mechanical experience, that perhaps energy and momentum are conserved only
statistically. By introducing neutrinos in 1930~\cite{Pauli1930} Pauli has saved the 
principle of energy and momentum conservation.

In 1934 the idea of Pauli led  Fermi~\cite{Fermi1934}  to the formulation of the theory of
nuclei $\beta$ decay, and generally to the theory of weak interaction. Fermi used the analogy
of electromagnetic interaction. It was only the effective theory, but now it is considered
as the early beginning of the modern gauge theory of weak interaction.  Even now the Fermi
theory is used to describe four fermion processes for small energy.

Knowing the neutron lifetime, Fermi was able to predict the value of the so-called
{\it Fermi constant}. Then it was possible to calculate the cross-section for inverse
$\beta$ decay and to predict a probability  of neutrino detection.  This probability was
very small, and  Bethe and Peierls claimed~\cite{BethePeierls} that neutrinos might
never be observed. Now this strong statement is a warning that in physics such ultimate opinions 
should not be stated. 22 years later electron neutrinos had been observed.

In 1937 the second charged lepton --- a  muon was discovered~\cite{NeddermeyerAnderson}.
Muons had very similar properties to almost 200 times lighter electrons. This similarity
led in the future to the concept of {\it lepton universality}.

As we want to be chronological, next we have to mention the Majorana idea of chargeless
fermion which are their own antiparticles --- and presently are known as {\it Majorana
particles}, even if this idea became popular only in the seventies.  In 1937 Majorana~\cite{Majorana1937}, 
using neutrinos, has suggested the existence of such elementary
objects.  Majorana spinors, as well as the older idea of Weyl spinors~\cite{Weyl1929}, was
connected with the space and charge parity non-conservation\footnote{The Weyl spinors do not
preserve the C and P symmetries. In the case of Majorana spinors it is not so. As was
noted later,  using these spinors,  it is possible to construct a theory, which satisfies
these two discrete symmetries.}. At that time it was unthinkable to accept breaking of
these two symmetries, and both ideas have been rejected.

In 1947, after discovery of muon decay, B.~Pontecorvo proposed the {\it universality}
of the Fermi interaction, at that time electron and muon. Then this
suggestion was
widely discussed in~\cite{universality} and possibly the origin of the current
{\it generation of leptons or a family} should be linked to this discussion.

In order to explain certain missing decay modes, in 1953 the concept of lepton number ($L$)
was introduced~\cite{Lconservation}. This is one of well known tested conservation law.
In the Fermi theory, in the Standard Model and, what is most important, in all present
experiments, $L$ is conserved.

Three years later first neutrino --- the {\it electron neutrino} has been
experimentally discovered in the inverse $\beta$ decay process. After considering several
methods, also a possible atomic bomb explosion, Reines and Cowan found
antineutrinos from a nuclear reactor~\cite{ReinesCowan}. It was the first experiment
which we call now {\it reactor neutrino experiment} where $\overline{\nu}_{e}$
produced by reactors were used.
\newpage

In 1956 neutrinos were used by Lee and Yang~\cite{LeeandYang1956} to put forward the
hypothesis, that the symmetries P and C are not satisfied in nature, and a year later~\cite{Wu1957}, 
to the experimental verification of this fact.

When it became apparent that the parity is broken, the weak Lagrangian has become much more
complicated and  scalar, vector and tensor terms together with parity violating  couplings
have to be taken into account. Such complicated situation was simplified  in 1958 when the
V--A theory of weak $\beta$ decay has been formulated~\cite{V-A}. Then the Weyl idea of two
component spinors, which describe massless fermion, finally found an application~\cite{masslessneutrino}, 
but even then it was noted that there is no difference between
Weyl and massless Majorana neutrinos~\cite{WeylMajorana}. The Weyl two component spinor
was simpler and there was no reason to use Majorana bispinors.

In 1958 the polarization of a neutrino has been measured in electron capture reaction 
$e^{-} + ^{152}$Eu$ \rightarrow ^{152}$Sm$^{*} + \nu_{e}$ and subsequent decay
$^{152}{\rm Sm}^{*} \rightarrow\break  ^{152}{\rm Sm} + \gamma$~\cite{neutrinohelicity}. The helicity of
neutrinos was negative in full agreement with two component theory of massless neutrino.

At the end of the fifties occurred one more thing that is sure to be noted. The concept of
{\it neutrino oscillation} was proposed by Pontecorvo~\cite{Pontecorvo1, Pontecorvo2}. 
Motivated by the $K^{0}\Leftrightarrow \overline{K}^{0}$ oscillation phenomena proposed 
by M.~Gell-Mann and A.~Pais in 1955, Pontecorvo suggested that similar phenomenon, transition between
$\nu\Leftrightarrow \overline{\nu}$  for Majorana neutrinos, can occur. He has interpreted
the (wrong) result of the Davis observation of $\overline{\nu} + ^{37}\!$Cl $\rightarrow e^{-}
+ ^{37}\!$Ar~\cite{Davisantineutrino} as a result of  $\overline{\nu} \Leftrightarrow \nu$
transition and then proper electron production $\nu + ^{37}\!$Cl $\rightarrow e^{-} +
^{37}\!$Ar.

One other thing happened in 1958 which have meaning for future discoveries. Feinberg
has tried to find muon decay $\mu\rightarrow e+\gamma$  without success~\cite{Feinberg}.

The fifties were indeed very fruitful for neutrino physics. At the end of this period we
had information that there are two charged leptons, electron and muon and one 
neutral --- electron neutrino. Lepton number $L$, which distinguishes leptons ($e^{-},\mu^{-},\nu_{e}$)
from antileptons ($e^{+},\mu^{+},\overline{\nu}_{e}$) was introduced, so all leptons were
Dirac particles. The nucleon  $\beta$ decay was described by V--A vector Lagrangian\vspace{-1mm}
\begin{eqnarray} \label{V-A}
{\cal L}_{\rm (V-A)} = \frac{G_{\rm F}}{\sqrt{2}}  \left(
\overline{\nu}_{e} \gamma^{\mu} \left(1-\gamma_{5}\right) e \right) \left(
\overline{n} \gamma_{\mu} (C_{\rm V}-C_{\rm A}\gamma_{5})p\right)   + {\rm h.c.}
\end{eqnarray}

\vspace{-1mm}
Neutrinos were massless particles described by Weyl two component spinors and took part in
the C and P violating, but CP conserving interaction~(\ref{V-A}).  The idea of neutrino
oscillation has appeared but not in the correct way, as a neutrino--antineutrino
transition.

In the next section we present the origin of the SM, which emerged in the sixties, and
the role of neutrinos in that time. In  Section 3 neutrino properties in the SM are presented
and the first experimental indications, which show that the theory must be extended
because the neutrinos do not satisfy the SM predictions and are massive particles. Then,
in Section 4, we describe, how the SM has to be minimally extended to predict the massive
neutrinos, how the phenomenon of neutrino oscillation is now understood and what kind of
experimental information about the neutrino masses and mixing we have today. In Section 5 we
present what are the consequences of the observed neutrino properties for physics beyond
the SM, and finally in Section 6 some conclusions are given.

\section{The road to the Standard Model}
\vspace{3mm}
The productive period in neutrino physics persisted also in the sixties. Already at the
beginning, as in many previous cases, Pontecorvo had a brilliant intuition and  
suggested~\cite{Pontecorvo3} that, if neutrino produced in the  pion decay $\pi^{+}
\rightarrow\mu^{+} + \nu_{\mu}$ cannot induce $e^{-}$, then both neutrinos $\nu_{e}$ and
$\nu_{\mu}$ are different particles. Such experiment was done in 1962 at Brookhaven
National Laboratory by L.M.~Lederman, M.~Schwartz,
\mbox{J.~Steinberger~{\it et al.} \cite{LSS}.}\break
Indeed, neutrinos from pion decay  always produced muons but never electrons. Then it was
clear that $\nu_{e}$ and $\nu_{\mu}$ are different particles and there are at least two
different family of leptons, a new neutrino, the {\it muon neutrino} appeared in particle
physics. It is also worth to stress that this Brookhaven experiment was in fact the
first, where the beam of neutrinos has been prepared, so it was the first experiment which
we now know as {\it accelerator neutrino experiment}.

In order to explain the smallness of leptonic decay of hyperons and a subtle difference of the
Fermi coupling $G_{\nu}$s between $\mu$ and $\beta$ decay, Maki, Nakagawa and Sakata (MNS) 
introduced neutrino mixing~\cite{MNS}. They assumed  that $\nu_{e}$ and
$\nu_{\mu}$ are not mass eigenstates, but are superposition of two neutrinos with
different masses
\begin{eqnarray} \label{MNS}
\nu_{e} &=& \nu_{1} \cos \theta  + \nu_{2} \sin \theta\,,\nonumber \\
\nu_{\mu} &=& -\nu_{1} \sin \theta  + \nu_{2} \cos \theta\,. 
\end{eqnarray}

In the MNS paper there was no discussion about neutrino oscillation. The first intuitive
understanding of neutrino mixing and oscillation was presented by Pontecorvo~\cite{Pontecorvo4} 
and by Gribov and Pontecorvo~\cite{GribovPontecorvo}. The full
theory of neutrino oscillation together with a third generation of leptons was finally
developed in 1975--76. The full $3\times 3$ mixing matrix appeared at that time and received the name
of the  Maki--Nakagawa--Sakata--Pontecorvo (MNSP) mixing matrix. Than it
has become possible the violation of CP symmetry in the lepton sector,
in the same way as for quarks.

In 1968 the Homestake solar neutrino experiment has started to work~\cite{Davis}.
Originally this experiment, by observing neutrinos produced in the sun,  had to check the
Bethe model~\cite{Bethe} for the creation of the solar energy. Nobody predicted that this
experiment will have to change its role and start to examine the properties of neutrinos.

In the sixties neutrinos, together with charged leptons and quarks, gave rise to the
formulation of the  model of electroweak interactions~\cite{WeinbergSalamGlashow}. At that
time only three quarks $(u,d,s)$ and four leptons $(e,\nu_{e}, \mu, \nu_{\mu})$ were known.
All neutrino properties known from the contact Fermi model have been preserved.  So,
neutrinos remain massless Weyl particles, which  interact only through the left-handed
currents, and  their interaction break up the discrete symmetries C and P maximally. There
is also a new property, not known before --- family leptons numbers $L_{e}, L_{\mu},
L_{\tau}$ which now are often used as {\it flavour lepton numbers}.

Over the next years to date, all the components of this model, with one exception, have
been  experimentally discovered. Three successive \mbox{quarks$\!$\cite{quarks}}, another charged
lepton $\tau$~\cite{Perl} and its neutrino $\nu_{\tau}$~\cite{DONAT} were found, but the
basic particle of the model --- the Higgs particle is still missing. Together with strong
interactions the full theory, which now describe all elementary particles interactions, is
known as the {\it Standard Model} (SM). In the model there are three families.  Four
leptons, known before and the new one, form the three families which are
called: electron, muon and tau families
\begin{equation}\label{families of leptons}
(\nu_{e}, e^{-})\,, \qquad   (\nu_{\mu}, \mu^{-})\,,  \qquad  (\nu_{\tau}, \tau^{-})\,.
\end{equation}

In this model neutrinos couple to charged gauge bosons $W^{\pm}$
\begin{equation}\label{charged current}
L_{\rm CC} = \frac{e}{2 \sqrt{2} \sin\theta_W}
\sum_{\alpha=e,\mu,\tau}\bar{\nu}_{\alpha}\gamma^{\mu}(1-\gamma_{5})l_{\alpha}W_{\mu
}^{+} + {\rm h.c.}\,,
\end{equation}
and to neutral $Z_{0}$ one
\begin{equation}\label{neutral current}
L_{\rm NC} =
\frac{e}{4 \sin\theta _W \cos\theta _W}\sum_{\alpha=e,\mu,\tau}
\bar{\nu}_{\alpha}\gamma^{\mu}(1-\gamma_{5})\nu_{\alpha}Z_{\mu}\,.
\end{equation}
Neutrino fields do not appear in any other place of the SM interaction\break Lagrangian.
From~(\ref{charged current}) and (\ref{neutral current})  we can easily find that really
the family lepton numbers $ L_{e}, L_{\mu}$ and $L_{\tau}$ are separately conserved and,
as a consequence, also the total lepton number $L = L_{e} + L_{\mu} + L_{\tau}$ is
conserved. In the SM there is no flavour lepton mixing ($L_{\alpha}$ conservation) and no
CP symmetry violation\footnote{Both properties, the lack of flavour mixing and the CP
symmetry conservation are connected with the assumption that neutrino are massless.}  in
the lepton sector, neutrinos are stable and have no electromagnetic structure (only the
charge radius $\langle r^{2} \rangle \neq 0 $).

There are three facts that determine that neutrinos are massless in
the~SM:
\begin{enumerate}
\item  we do not introduce the right-handed fields $\nu_{\rm R}$, 
\item in the model only one Higgs doublet is introduced, 
\item we require that the theory is  renormalizable. 
\end{enumerate}
As we see none of these reasons is very basic. Lack of mass of neutrinos is not
guaranteed by any fundamental theory so resignation from any of the previous conditions
1--3 results in a mathematically correct theory. So it is very easy to find a theory
with massive neutrinos, the problem is, as we will see, why their masses are so remarkably
small.

  Neutrinos are responsible for the first great success of the SM. In 1973, the
measurement  of  a neutral current reaction in the Gargamelle bubble 
chamber~\cite{Gargamelle1973} experiment at CERN has given the first indication that the neutral
gauge boson $Z_{0}$ exists. Neutrinos from pion decay scattered on the liquid 
scintillator --- freon, and muon has not been produced $\nu_{\mu} + N \rightarrow  \nu_{\mu} + N$
After this experiment, SM received a good foundation even if successive quarks and
leptons have been discovered  later. The bases of the model have survived to this day,
only the number of generations has grown.  So far there are no  experimental facts which
are incompatible with it, with one exception. From many different experiments it is now
obvious that neutrinos are not massless particles.

\section{Neutrinos in the SM and the road outside}

In the past 40 years MS has achieved a great success and now is firmly established as the
model for lepton and quarks interactions. In all experiments, where the particles,
including the neutrinos, collide or decay,  the energy is much larger than the masses of
neutrinos and the correspondence between theory and data is very good or good.
Neutrinos also contributed to this success. The measurements in LEP in 1989 of the 
so-called invisible $Z_{0}$ boson decay width, interpreted as  the  $Z_{0}$ decays into
unobserved neutrinos, has fixed the number of generations at three~\cite{Zdecaywidth}.
Moreover, this result excluded the existence of other neutrinos, which couple to $Z_{0}$
and have mass smaller than  half of the $Z_{0}$ boson mass.

Since the first experiment of Reines and Cowan~\cite{ReinesCowan}, where the electron
antineutrino $\bar{\nu_{e}}$  was discovered, neutrinos have been observed continuously in many
different processes. The cross-sections for (anti)neutrino + electron, (anti)neutrino +
nucleon and (anti)neutrinos + nuclei  are measured at different neutrino energies and for
different final channels. All measured cross-sections agree with the SM predictions with
massless neutrinos, so only upper limit for the neutrino masses can be found. The best
upper limit for the neutrino effective mass ($m_{i}$ are neutrino masses and $U_{ei}$ are
the elements of the mixing matrix,  see the next section)~\cite{effectivemass}
\begin{equation}\label{mass limit}
m_{\beta} = \sqrt{\sum_{i= 1,2,3} \vert U_{ei}\vert ^{2} m_{i}^{2}}\,, \qquad   m_{\rm min} <
m_{\beta} < m_{\rm max}\,,
\end{equation}
has been found in a tritium $\beta$ decay~\cite{betadecay}
\begin{equation}\label{value of mass limit}
m_{\beta} < 2.2~{\rm eV}\,.
\end{equation}
The problems of the SM began with the previously mentioned experiment in  Homestake where
Davis in 1968 installed the detector in order to catch neutrinos produced inside the
Sun~\cite{Davis}. Just from the beginning it was observed that number of neutrinos
detected in the Chlorine detector in Homestake was only one third of the one predicted by the 
so-called Standard Solar Model~\cite{Bahcall}. This discrepancy between the number of
predicted neutrinos and the number measured in this first solar neutrino
experiment was
known as {\it The Solar Neutrino Problem}.
Many physicists have not believed that the problem can be solved by changing the SM, but
rather that the solution to the problem lies with a wrong neutrino flux  given by the SSM,
or a misinterpretation of the experiment. However, collected over many years the results
from Homestake~\cite{Homestake}  and all subsequent experiments of KAMIOKANDE
Collaboration~\cite{KAMIOKANDE}, SAGE Collaboration~\cite{SAGE}  and GALLEX 
Collaboration~\cite{GALLEX} confirmed the first results of Davis. It is also worth to stress, that the
water detector in Kamioka mine in Japan was not built to study neutrinos, but to look for
proton decay. With time  the main objective has been changed  and Kamiokande and later
SuperKamiokande became the most important neutrino detector which looked for the neutrino
flavour change.

In the eighties the problem with neutrinos has become wider. Similar alarming phenomenon --- the
flavour change of neutrinos  in the atmospheric neutrino flux --- was observed. Several
experimental groups have started to observe atmospheric neutrinos deficit~\cite{KAMIOKANDE2,
IMB,Soudan2} but not all~\cite{NUSEX, Frejus} (for a review see~\cite{AtmosphericReview}).
After these experimental facts the situation was still not clear.  The problem was
resolved at the end of nineties. In 1998  the phenomenon of neutrino oscillation was definitively 
confirmed~\cite{SuperKamiokande}, showing that neutrinos have mass. This was the first evident
indication, that the SM has to be extended. Finally, in 2002 the solar neutrino problem
was ultimately resolved~\cite{SNO}. The Sudbury Neutrino Observatory (SNO) Collaboration
made a unique measurement in which the total number of neutrinos (having energy above detector
threshold) of all types (not only electron neutrinos) was observed. The longest running
experiments, the solar neutrino experiments, have finished in 2002 after 34 years and have
given spectacular success of the Solar Standard Model. The SNO  together with the
SuperKamiokande measurements (for recent result see~\cite{SuperKamiokandeII} and
\cite{SNOnext}) show that most of the neutrinos produced in the interior of the Sun as electron neutrinos,
are changed into muon and tau neutrinos by the time they reach the Earth.  As predicted by
B.~Pontecorvo neutrinos oscillate. Presently the oscillation phenomena have been confirmed
by accelerator neutrino  K2K~\cite{K2K} and MINOS~\cite{MINOS} as well as reactor neutrino
experiments~\cite{reactor}. The last reactor experiment (KamLAND) was
very important. Its results combined with all the earlier solar neutrino
data established the correct parameters for the solar neutrino deficit.
Taking into account all data, especially the SuperKamiokande and
KamLAND, we have now very well evidence for neutrino disappearance and
reappearance and all non-oscillations models are eliminated.

\section{Neutrinos in the SM with small neutrino mass --- the $\nu$SM}

From various neutrino oscillation experiments it results  that the SM must be extended at
least in such a way, that neutrinos must be massive. There are many beyond the SM (BSM)
theories which satisfy such a requirement. The simplest and popular scenario is such that
the neutrino mass is the only one visible result of New Physics (NP) at very high scale
({\it e.g.} unification scale $\sim 10^{16}$~GeV), and all other BSM interaction of quarks and
charged leptons are  completely negligible at low, experimentally accessible energies.
Such model is sometimes called the New SM = $\nu$SM. In such model the NP is {\it visible}  by
the neutrino mass Lagrangian and neutrino mixing matrix. The mass term and mixing matrix
distinguish Dirac from Majorana neutrinos. In the case of Dirac neutrinos the mass term
has the form
\begin{equation}\label{Dirac mass term}
L_{\rm mass}(D) = \sum_{i=1,2,3} m_{i}^{D} \left(\bar{\nu}_{iR} \nu_{iL} + \bar{\nu}_{iL}
\nu_{iR}\right)\,.
\end{equation}
For Majorana neutrino two kinds of mass Lagrangian are allowed, built using the
left-handed chiral fields
\begin{equation}\label{Majorana1 mass term}
L_{\rm mass}^{\rm L}(M) = \sum_{i=1,2,3} m_{Li}^{M} \left(\bar{\nu}_{iR}^{c} \nu_{iL} + \bar{\nu}_{iL}
\nu_{iR}^{c}\right)\,, \qquad \nu_{iR}^{c}= i \gamma^{2} \nu_{iL}^{*}
\end{equation}
and the right-handed fields
\begin{equation}\label{Majorana2 mass term}
L_{\rm mass}^{\rm R}(M) = \sum_{i=1,2,3} m_{Ri}^{M} \left(\bar{\nu}_{iL}^{c} \nu_{iR} + \bar{\nu}_{iR}
\nu_{iL}^{c}\right)\,, \qquad \nu_{iL}^{c}= i \gamma^{2} \nu_{iR}^{*}\,.
\end{equation}
In the $\nu$SM the charged~(\ref{charged current}) and neutral~(\ref{neutral current})
current Lagrangians have a new form
\begin{equation}\label{new charged current}
L_{\rm CC}^{\nu {\rm SM}} = \frac{e}{2 \sqrt{2} \sin\theta _W}\sum_{\alpha,
i}\bar{\nu}_{i}\gamma^{\mu}(1-\gamma_{5}) U_{\alpha i}^{*} l_{\alpha}
W_{\mu}^{+} + {\rm h.c.}
\end{equation}
and for $Z_{0}$
\begin{equation}\label{new neutral current}
L_{\rm NC}^{\nu {\rm SM}} =
\frac{e}{4 \sin\theta _W \cos\theta _W}\sum_{i=1,2,3}
\bar{\nu}_{i}\gamma^{\mu}(1-\gamma_{5})\nu_{i}Z_{\mu}\,.
\end{equation}
As now neutrinos are massive particles the interaction  with neutral
Higgs particle appears\vspace{-3mm}
\begin{equation}\label{with Higgs}
L^{\nu {\rm SM}}_{H} = \frac{e}{2 \sin\theta _W}\sum_{i=1,2,3} \left(\frac{m_{i}}{M_{W}}\right)
\bar{\nu}_{i} \nu_{i}H\,,
\end{equation}
but the ratio $\frac{m_{i}}{M_{W}}\ll1$, and  the neutrinos coupling to Higgs particles
(Eq.~(\ref{with Higgs})) is negligible small. In such models the lepton flavour numbers
are not conserved separately, and neutrinos can oscillate.  The total lepton number is
(Dirac neutrinos) or is not (Majorana neutrino) conserved. The CP symmetry is  broken if
the complex phases in the U mixing matrix are different than $\delta_{\rm CP}, \phi_{1}, \phi_{2}
\neq 0, \frac{\pi}{2}, \pi$ (for the present parametrisation of the MNSP mixing matrix see Ref.~\cite{Nakamura}).

The full theory of neutrino oscillation was worked out in several papers~\cite{Oscillationtheory},
although as we can see from the amount of publications that are constantly
emerging, it is still a moot (see {\it e.g.}~\cite{OscillationtheoryII}). Without going into
details\footnote{The full description of the oscillation process need the use of the wave
packet for all particles which appear in the neutrino production and detection process,
see {\it e.g.}~\cite{book}.}, we can understand the neutrino oscillation as a typical phenomenon
of relativistic quantum mechanics, in which the states of particles with different masses
may be added in a coherent way and interfere\footnote{In the nonrelativistic quantum
mechanics, particles of different masses belong to separate Hilbert spaces, and do not
interfere.}. This in short can be summarized in the following way:
\begin{itemize}\parskip2pt
\item[\it (i)] In each production process, which in the lowest order is described by the CC
Lagrangian~(\ref{new charged current}),  neutrinos with different masses ($m_{i}$) are
produced.
\item[\it (ii)] In any realistic production process the mass differences $\vert m_{i} -
m_{k}\vert$ for any two produced neutrinos is much smaller than the neutrino mass
uncertainty, determined from measured energies and momenta of all particles in the
production/detection process without neutrinos\vspace{-2mm}
\begin{equation}\label{condition for oscillation}
\vert m_{i} - m_{k}\vert \ll \Delta m \equiv \frac{E \Delta E+p \Delta p}{\sqrt{E^{2} -
p^{2}}}\,,
\end{equation}
where $E, (\Delta E)$ and $p, (\Delta p)$ are energy and momentum of neutrino and their
uncertainties.
\item[\it (iii)] From the Lagrangian~(\ref{new charged current}) it follows, that any flavour
state $\alpha = e, \mu, \tau$ of  produced or detected neutrinos\footnote{The neutrino
flavour is determined by the charged leptons, which  appear in the production or detection
processes.}  is the linear combination of the mass states

\begin{eqnarray}
\label{neutrin SM states}
\vert\nu_{\alpha},\downarrow\rangle
=\sum_{i}U_{\alpha,i}^{*}\vert\nu_{i},\downarrow\rangle\,, \qquad
\vert\overline{\nu}_{\alpha},\uparrow\rangle
=\sum_{i}U_{\alpha,i}\vert\overline{\nu}_{i},\uparrow\rangle\,,
\end{eqnarray}
where  the arrows $(\downarrow\uparrow)$ denote the helicities of neutrino (antineutrino)
which all the time in the oscillation process do not change.
\item[\it (iv)] If neutrinos $\alpha$ are produced in the production point,  and placed at a
distance $L$ a detector is looking for $\beta$ flavour neutrinos, then the amplitude for
flavour $\alpha \rightarrow \beta$ change is given by\vspace{-1.5mm}
\begin{eqnarray} \label{amplitude for flavour change}
A_{\alpha \rightarrow \beta} (E,L) = \left\langle \nu_{\beta} \left\vert e^{-i H t} \right\vert
\nu_{\alpha}\right\rangle\,,
\end{eqnarray}
where $H$ is the Hamiltonian which describes neutrino propagation in vacuum or in 
matter\footnote{The presented way of finding the amplitudes works only in frame of the $\nu {\rm SM}$.
If in a neutrino production and/or detection processes beyond the SM interactions play a
role, a more complicated formalism for neutrino oscillation have to  be used 
(see {\it e.g.}~\cite{OscillationtheoryIII}).}.
\item[\it (v)]  Then the  probability of the neutrino flavour change after propagation of
distance $L$ in the vacuum, is  given by
\vspace{-1.5mm}
\begin{eqnarray} \label{probability for flavour change}
P_{\alpha \rightarrow \beta}(E,L)  &=&   
\vert A_{\alpha \rightarrow \beta} (E,L) \vert^{2}  \nonumber \\
&=&\left\vert \sum_{i} \sum_{k} U_{\beta i} U_{\alpha k}^{*} \left\langle \nu_{i} \left\vert
e^{-i\sqrt{E_{i}^{2} + p_{i}^{2}}  \frac{L}{v_{i}}} \right\vert \nu_{k}\right\rangle \right\vert^{2} \nonumber \\  
&=&\sum_{i} \sum_{k} U_{\alpha i} U_{\beta k} U_{\alpha k}^{*} U_{\beta i}^{*}
e^{i\frac{\delta m_{i k}^{2} L}{2 E}},
\end{eqnarray}
where $\delta m_{i k}^{2} =m_{i}^{2} - m_{k}^{2} $ and $E$ is the average energy of
neutrinos.
\end{itemize}

We see why it was so difficult to get any information that neutrinos are massive
particles. Independently how small the difference of a neutrino mass square ($\delta m_{i
k}^{2}$) is, the other factor $L/E$ depends on our choice and can be large,
such that the total phase $(\delta m_{i k}^{2} L/2 E)$ is large too, and the
effect of  neutrino oscillation can be visible.
In any laboratory experiment the neutrino detection cross-section {\it e.g.} on electron
$\sigma(\nu_{e} + e^{-} \rightarrow
\nu_{e}+ e^{-} )\break \approx 9.5 \times 10^{-49} \left(\frac{E_{\nu}}{1~{\rm MeV}}\right)~{\rm m}^{2} $ or for
inverse $\beta$ decay process $\sigma(\nu_{e} + n \rightarrow
e^{-}+ p )\break \approx 9.3 \times 10^{-48} \left(\frac{E_{\nu}}{1~{\rm MeV}}\right)~{\rm m}^{2}$ is very small and
proportional to the neutrino energy in Lab system. So practically detected neutrinos are
relativistic,\break   $(m_{\nu}/E_{\nu}) \rightarrow 0$.
Therefore, in any laboratory process where neutrinos are observed, neutrino masses can be
neglected. The observed  family  lepton numbers $L_{\alpha}$  conservations follow from
unitarity of the MNSP mixing matrix.

Also the so-called {\it confusion theorem} was proven~\cite{DMconfusiontheorem},
which states that differences in all observables for Dirac and Majorana neutrinos  due to
the different mass Lagrangians~(\ref{Dirac mass term}), (\ref{Majorana1 mass term}),
(\ref{Majorana2 mass term}) smoothly disappear for $m_{i}\rightarrow 0 $.

From present experimental data we have information about neutrino masses and about the
elements of the MNSP mixing matrix.
Direct information about neutrino masses come from the tritium $\beta$~decay and are given
by Eq.~(\ref{value of mass limit}).
Some information also come from a neutrinoless double $\beta$~decay~\cite{neutrinoless},
which can occur only if neutrino are Majorana particles, and if such decay is observed, it
is possible to measure the other effective neutrino mass
\begin{equation}\label{effective bb mass}
\langle m_{0\nu} \rangle= \left\vert\sum_{i= 1,2,3} U_{ei} ^{2} m_{i}\right\vert\,, \qquad   \langle
m_{0\nu} \rangle< m_{\rm max}\,.
\end{equation}
Latest experimental results from the CUORICINO~\cite{CUORICINO} give
\begin{equation}\label{effective bb mass limit}
\langle m_{0\nu} \rangle < 0.19-0.68~{\rm eV} \Longrightarrow m_{\rm max} >  0.68~{\rm eV}\,,
\end{equation}
from which we can conclude that the mass of the heaviest neutrino $m_{\rm max} > 0.68~{\rm eV}$.
These results have very large systematic error which emerges from large discrepancy
between different nuclear matrix element calculations\footnote{CUORICINO experiment
measure the decay timelife of $^{130}$Te and  they found $T^{0 \nu}_{1/2} > 3.0 \times
10^{34}\, {\rm y}\,(90$\%$~{\rm C.L.})$.}.
From neutrino oscillation experiments we also know  two differences of neutrino masses
squared. The last global fits~\cite{globalfitsI} give
\begin{equation}\label{delta m12 square}
\delta m_{21}^{2} = m_{2}^{2} - m_{1}^{2}  = (7.05 - 8.34) \times
10^{-5}~{\rm eV}^{2}\,,
\end{equation}
\begin{equation}\label{delta m31 square}
\left\vert \delta m_{31}^{2}\right\vert = \left\vert m_{3}^{2} - m_{1}^{2}\right\vert  = (2.07 -
2.75) \times 10^{-3}~{\rm eV}^{2} \Longrightarrow m_{\rm max} >  0.045~{\rm eV}\,.
\end{equation}
From~(\ref{delta m31 square}) it follows that the mass of the heaviest neutrino must be
$m_{\rm max} > 0.045$~eV.

The elements of the unitary mixing matrix $U_{\alpha i}$ are parametrized by the three
angles  $\theta_{12}, \theta_{23}, \theta_{13} $  and one CP violating phase $\delta$.
Currently not all these parameters are known.  Combined data give~\cite{globalfitsI}
(with $3\sigma$ interval)
\begin{equation}\label{values of theta angles}
\sin^{2}\theta_{12} \in (0.25 - 0.37)\,, \quad  \sin^{2}\theta_{23} \in
(0.36 - 0.67)\,, \quad \sin^{2}\theta_{13} < 0.056\,.
\end{equation}

The result of atmospheric, solar, reactor (KamLAND) and accelerator (K2K and MINOS)
neutrino experiments are very well explained by the neutrino oscillations in the framework
of the three neutrino mixing. We have a rather precise knowledge of the values of
squared-mass difference $\delta m^{2}_{21}$, the absolute value of $\vert \delta
m^{2}_{31} \vert$ and the values of two mixing angles $\theta_{12}$  and  $\theta_{23}$.
We expect that the next generation of different experiments will give us information
about: {\it (i)} the absolute scale of neutrino mass connected with the spectrum of
masses (normal hierarchy, inverted hierarchy, degenerate)\footnote{Up to now we  have
information that heaviest neutrino mass is in the range $0.05$~eV ($0.68$~eV)$<
m_{\rm max} < 2.2$~eV.},  {\it (ii)} nature of neutrinos (are they Dirac or Majorana
particles),\break {\it (iii)} value of  $ \theta _{13} $ mixing angle (is  $\theta_{13}$
close to zero or rather close to the upper limit)\footnote{Last data including the
results from Borexino experiment~\cite{BOREXINO} found non zero value for
$\theta_{13},\sin^{2} \theta_{ 13} = 0.0095^{ +0.013}_ {0.007}$~\cite{globalfitsII}.}
and finally, {\it (iv)} the CP violating phases ($\delta_{\rm CP}$ --- the only one for
Dirac neutrinos, or additional two, $\phi_{1}, \phi_{2}$ for Majorana neutrinos). It is
worth paying attention to the fact that the masses of some neutrinos can be smaller then
the experimental error of the charged lepton masses (for electron $(\Delta m_{e})_{\rm exp}=
0.013$~eV).
\section{Beyond the SM}

The very small mass of neutrinos  and the completely different leptonic mixing matrix in
comparison to quarks require a modification  of the SM and some  New Physics (NP)
beyond the SM must be found. Unfortunately, data are not precise enough  to indicate which
NP model should be chosen. In the previous section we have considered the NP at the
unification scale $(10^{15}$~GeV), now we will concentrate on a NP which appears in the
present available energy scale. There are many hints that really NP operates at a 0~(TeV)
scale~\cite{NP}. Such NP is much more interesting, there is a chance to discover it at the
LHC, and next high energy machines ({\it e.g.} ILC) or at the future more precise neutrino
experiments. Problem of neutrino mass and mixing can refer to the unification scale as
well as to the 0~(TeV) scale. If a NP modifies the neutrino interaction at 0~(TeV) scale
then, as we mentioned before, the description of oscillation must be modified too. We shortly
discuss ourselves to that.
\subsection{ Neutrino mass and mixing}

Neutrino masses are  much smaller than the masses of charged leptons and quarks. For
mixing angles it is opposite, there are two large mixing angles for leptons which
contrast sharply with the smallness of the quark mixing angles. We would like to know why
it is so. On the other hand, the problem of particle masses waits for a solution. Why do we try to 
solve separately the neutrino mass and the flavour problem? The ratio of the
electron mass to neutrino masses  $\frac{m_{\nu}}{m_{e}}\leq 10^{-6}$ is almost the same
as the ratio of the top quark to electron  $\frac{m_{e}}{m_{t}}\simeq 10^{-5}$. There are 
several reasons why the smallness of neutrinos masses is interesting.
Firstly, the smallness of neutrino mass remains a question even within one family. Quark
mass ratio in the same family is about 10, while for the same lepton generation the mass
ratio is smaller than $10^{-6}$. Secondly, the problem of neutrino mass may be connected
with their nature. The quarks and charged leptons are Dirac particles.  Neutrinos have
probably a Majorana nature. And finally, even if the problem of mass is not resolved,
the large difference for lepton masses within a single family and completely different
structure of the mixing matrix\footnote{For the quark mixing, the non-diagonal elements of
the CKM mixing matrix are very small, which is presumably due to small ratios of the quark
masses $m_{c}/m_{t}, m_{u}/m_{c}, m_{s}/m_{b}, m_{d}/m_{s}$, 
such relations between elements of the MNSP matrix and the ratios
of neutrino masses  do not exist.} can shed a light on the extension of the SM. This is
probably the main reason why the problem of neutrino mass, usually connected with the
flavour problem,  is so intensively studied in recent years (see {\it e.g.}~\cite{Massproblempapers}, 
for more complete list see~\cite{MZ}).

The simplest way to get massive neutrinos is to add to the SM fields $N$ right-handed
chiral neutrino fields ($\nu_{\beta R}, \beta=1,2, \dots N$) and to introduce the neutrino
masses in the same way as for the quarks and charged leptons
\begin{equation}\label{Simplest way to get mass}
L_{Y} = - \sum_{\alpha,\beta} f_{\alpha,\beta} \tilde{\psi}_{\alpha
L}(-i\sigma_{2}\varphi^{*})\nu_{\beta R} + {\rm h.c.}\,,
\end{equation}
where $\tilde{\psi}_{\alpha L}$ and $\varphi$ are SU(2) doublets fields of leptons and
Higgs particles respectively.
There is no fundamental reason why we cannot do that, but we do not like this solution.
Neutrino mass matrix is proportional to the Yukawa couplings $f_{\alpha,\beta}$ and there
is no good reason why these couplings must be so small. Such a solution does not give any
indication how to extend the SM.

The other possibility is to add to the previous model the right-handed mass term
\begin{equation}\label{Simplest way with RH}
L_{\rm RH} = -\frac{1}{2} \sum_{\alpha,\beta} g_{\alpha,\beta} \tilde{\nu}_{\alpha L}^{c}
\nu_{\beta R} + {\rm h.c.}
\end{equation}
Now we have three possibilities.  The most popular is the so-called {\it see--saw
mechanism}. The $g_{\alpha,\beta}$ Yukawa constants are very large ($\vert
g_{\alpha,\beta} \vert \gg \vert f_{\alpha,\beta} \vert$), then for $N=3$ we can get three
light and three heavy Majorana particles and B--L symmetry is broken. As usually, if two
very different scales exist, we meet with the hierarchy problem. If the large scale has a
quantum gravity range, neutrinos obtain too small masses, $m\sim 10^{-5}$~eV. The next
possibility is the case where  $g_{\alpha,\beta}$ are very small ($\vert g_{\alpha,\beta}
\vert \ll \vert f_{\alpha,\beta} \vert$), then the so-called {\it pseudo-Dirac
neutrino} scenario is realized~\cite{PseudoDirac}. The neutrinos are almost Dirac
particles with very tiny amount of the Majorana mass. It was found that then the Yukawa
coupling must be very small in order to be consistent with current solar neutrino
observation~\cite{smallYukawa}. Recently, the third possibility was considered where the
Yukawa constants  $f_{\alpha,\beta}$ in~(\ref{Simplest way to get mass}) and
$g_{\alpha,\beta}$ are of the same order ($\vert g_{\alpha,\beta} \vert \simeq \vert
f_{\alpha,\beta} \vert$)~\cite{Schizophrenicneutrinos}, then some mass states can be
Dirac and the others Majorana. The flavour neutrinos which are combination of two Dirac
and one Majorana or one Dirac and two Majorana neutrinos was called {\it schizophrenic neutrinos}.

In the way presented up to now we were able to give mass to neutrinos without a systematic
knowledge on how the SM must be extended.  There are a lot of various models which in a
better or  worse way explain  small neutrino masses and large two mixing angles. The first
option is to continue to maintain symmetry of the SM and, {\it (i)} modify the fermion
sector, {\it (ii)} enlarge the  Higgs sector, and {\it (iii)} break spontaneously the
B--L symmetry (Majoron(s) appears). The first possibility, as we have discussed previously,
was not satisfactory. There are three working ways of the Higgs sector enlargement, where
(1) additional Higgs triplet ${\mit \Delta}$, (2) singly charged singlet,  $h_{-}$, or (3) doubly
charged gauge singlet $k_{++}$ are introduced.  These possibilities are very popular. In
models with the Higgs triplet the see--saw mechanism is operating. Models with additional
singlets, invented by Zee and Babu~\cite{ZeeandBabu}, are very interesting as NP
appears at TeV scale and there is a chance to see some implication at LHC.  Neutrino
masses are small, as they are generated at either one or two loops.
There are also two different realizations of models with Majorons, (1) a gauge singlet
and additional right-handed neutrinos are introduced, or (2) the only Higgs sector is
extended by adding a Higgs triplet and a singly charged scalar.

The second option is to abandon the symmetry group of the SM and build a model which at
low-energy has all  features of the SM. Several such models are considered in the
literature, {\it (i)} new gauge group ${\rm SU}_{\rm L}(2)\otimes {\rm SU}_{\rm R}(2) \otimes {\rm U}(1)_{\rm B-L}$
with two Higgs doublets or Higgs doublets and triplets, {\it (ii)} models of grand
unification based on SU(5), SO(10) or $E_{6}$ symmetry group, {\it (iii)}
supersymmetric models in several versions, the MSSM, the model with broken R-parity  and
models based on the supersymmetric Left--Right group.

The next problem is connected with the specific structure of the flavour mixing matrix
the MNSP matrix. In order to understand  the large values of mixing angles
$\theta_{12},\theta_{23}$ and much smaller angle $\theta_{13}$, special flavour symmetry
are usually imposed in the models.

Despite some successes in understanding of the problem of the small neutrino masses and
the large mixing angles, it is difficult to accept that this problem is solved.  From the
case considered at the beginning of this section we see that so different scenarios of the
neutrino mass matrix still agree with experimental data (see--saw mechanism, pseudo-Dirac
neutrinos, schizophrenic neutrinos). This situation is probably connected with still
too poor experimental knowledge of the neutrino masses and mixing matrix elements and the
selection of the best theoretical model is difficult.

\subsection{Neutrino oscillation beyond the SM}

The original description of the neutrino oscillation phenomena, as we show before, was
introduced in  the mid seventies of the last century~\cite{Oscillationtheory}. Such description works
well in the case of $\nu {\rm SM}$ but does not work if the NP modify neutrino production and
detection processes.

Recently a full description has been  proposed, which may be used not only for the $\nu
{\rm SM}$ but can be applied for any model of neutrino interactions, and in which the neutrinos
propagate over long distances on mass shell~\cite{OscillationtheoryIII}. The neutrino
states are obtained from the dynamics of a production/detection process and the
entanglement between produced/detected particles is taken into account. This new approach
is presented shortly. As  a  production process the three body decay ({\it e.g.} muon  or
nuclear $\beta$ decay) is considered
\begin{eqnarray}
\label{process for neutrino production}
A \rightarrow B + \overline{l}_{\alpha} + \nu_{i}(\lambda)\,.
\end{eqnarray}
The state of produced neutrinos in this process, in the rest frame of the decaying
particle $A$, is described by the density matrix which depends on the dynamics of the 
process~(\ref{process for neutrino production}).  In the base where the neutrino mass $(m_{i})$
and helicity $\lambda$ are specified ($\vert\nu_{i},\lambda\rangle$), the density matrix
is given by the well known formula
\begin{eqnarray} \label{neutrino density matrix1}
&&\hspace{-4mm}\varrho ^{\alpha} (\lambda, i; \eta, k; E,\theta,\varphi) =\nonumber \\
&&\hspace{-4mm}\frac{1}{N_{\alpha}}\! \sum_{\rm spins}\!\int\!\overline{d {\rm Lips}} A^{\alpha}_{i}
\!\left(\lambda_{A};\lambda_{B},\lambda_{l},\lambda; E,\theta,\varphi\right)
\varrho_{\lambda_{A},\lambda_{A'}}  A^{\alpha *}_{k}
\left(\lambda_{A'};\lambda_{B},\lambda_{l},\eta;
E,\theta,\varphi\right)\,, \nonumber \\
\end{eqnarray}
where the integral $\overline{d{\rm Lips}}$ is taken over the part of the phase space, without
neutrinos energy ($E$) and its momentum direction $(\theta,\varphi)$,  the
$\varrho_{\lambda_{A}, \lambda_{A'}}$ is the density matrix which describes the
polarization of decaying particle ($A$) and the factor $N_{\alpha}$ normalizes the density
matrix, such that  ${\rm Tr} \varrho =1$.

Let us assume that in the detection process the lepton of flavour $\beta$ is produced in
our detector
\begin{eqnarray}
\label{neutrino detection process}
\nu_{i} + C \rightarrow l_{\beta} + D\,,
\end{eqnarray}
then the total cross-section for neutrino detection is calculated in the usual way
\begin{eqnarray}
\label{detection cross-section}
&&\hspace{-4mm}\sigma_{\alpha \rightarrow\beta} (E,L)\nonumber \\
&&\hspace{-4mm}=\frac{1}{64\pi ^2
s}\frac{p_{f}}{p_{i}}\frac{1}{2s_{C}+1} \sum_{\rm spins, masses}\int d{\mit \Omega}
\ f_{i}^\beta}(\lambda){ \varrho ^{\alpha}(L;i,\lambda;k,\eta) \   f_{k}^{\beta *}(\eta)\,,
\end{eqnarray}
where the $f_{i}^{\beta}(\lambda)$ are spin amplitudes for the detection 
process~(\ref{neutrino detection process}) of neutrino with mass $m_{i}$ and helicity $\lambda$.
The $\varrho^{\alpha}(L;i,\lambda;k,\eta)$ is the density matrix  after neutrino
propagation, calculated in the following way
\begin{eqnarray}
\label{time evolution}
\varrho ^{\alpha}(E, L) = e^{-iHL} \varrho ^{\alpha}(E, L=0)  e^{iHL}\,.
\end{eqnarray}
In such proposed approach, depending on the neutrino interaction in the production
process, the initial neutrino state can be pure, as in the $\nu$SM,  or mixed. The final
formula for the detection rate does or does not factorize to neutrino oscillation
probability times detection cross-section (for details see~\cite{OscillationtheoryIII}).

\section{Conclusions}

Some selected topics from the neutrino history have been described, mainly from the past
50 years. Neutrinos have always given  new and unexpected information about elementary
interaction. Neutrinos are very special because they only weakly interact. For this
reason, in a special way helped in the formulation of the theory of electroweak
interactions.  A few experiments, which were prepared for other purposes, after some time
have begun to explore properties of neutrinos.  In such, somewhat accidental way,
experiments began, which finally led to the discovery of neutrino mass. The disclosure that
neutrinos are massive particle is probably one of the most important discoveries in
particle physics in recent years. Although it did not change much in the laboratory
experiments, where very small neutrino mass does not play a role, the discovery is of
great importance for particle physics, astrophysics and cosmology.
In physics of elementary interaction, neutrinos has opened the window into phenomena
beyond the SM. In astrophysics, neutrinos allow ``glimpses of the interiors of stars'' and
to verify the theory of the processes taking place inside, {\it e.g.} verify the Standard
Solar Model or different models of supernova explosion. In cosmology, once we manage to
develop a detecting method of  relic neutrinos, which posses a very low-energy, neutrinos
will examine perhaps the evolution of the Universe in the first seconds after the Big
Bang. Heavy neutrinos may help to solve the riddle of Dark Matter, and understand the
problem of barion asymmetry via leptogenesis. To answer many of these questions better
information about neutrino properties are needed. Many of neutrino experiments are still
collecting data, new experiments are planned. There are many``working groups'' which discuss the new
experiments with very intensive beam of neutrinos (beta beams, superbeams, neutrino factories)
and larger and better detectors. The field is extremely active. It seems that next years will be very
interesting in the physics of neutrinos.

\vspace{7mm}
It is a great pleasure to thank the organizers of the L
Cracow School of Theoretical Physics for invitation and opportunity to present this
material. This work has been supported by the Polish Ministry of Science and Higher Education under grant No. N
N202 064936. The author would like to thank the colleagues from the Department of
Theoretical Physics and Cosmology at the University of Granada in Spain for the pleasant
atmosphere and hospitality during the stay, to F. del Aguila and A.
Bueno for many valuable remarks, and the Junta de Andaluc\'ia for support (FQM 03048).

\end{document}